\documentclass[11pt]{amsart}
\usepackage{latexsym,amssymb,amsmath,amscd,amsthm}
\numberwithin{equation}{section}
\theoremstyle{remark}

\newcommand{\bq}{\begin{equation}}
\newcommand{\bea}{\begin{array}}
\newcommand{\eea}{\end{array}}

\newcommand{\ga}{\alpha}
\newcommand{\gep}{\epsilon}
\newcommand{\gD}{\Delta}
\newcommand{\gl}{\lambda}
\newcommand{\gL}{\Lambda}
\newcommand{\gb}{\beta}
\newcommand{\ot}{\otimes}
\newcommand{\mf}{\mathfrak}
\newcommand{\mc}{\mathcal}

\newcommand{\dg}{\dagger}

\newcommand{\gO}{\Omega}
\newcommand{\gG}{\Gamma}
\newcommand{\gt}{\theta}
\newcommand{\gs}{\sigma}

\newcommand{\gag}{\gamma}
\newcommand{\gd}{\delta}
\newcommand{\pp}{\partial}

\newcommand{\tl}{\tilde}
\newcommand{\na}{\nabla}
\newcommand{\gk}{\kappa}

\newcommand{\bl}{\blacklozenge}
\newcommand{\bs}{\blacksquare}

\newcommand{\bgs}{\bigstar}

\newcommand{\gS}{\Sigma}

\newcommand{{\DDD}}{D\!\!\!\!\!\!-}

\title{ENTROPY, GEOMETRY, AND THE QUANTUM POTENTIAL}

\author{Robert Carroll\\University of Illinois, Urbana, IL 61801}

\date{November, 2005\thanks{email: rcarroll@math.uiuc.edu}}

\begin{document}

\bibliographystyle{plain}

\begin{abstract} 
We sketch and emphasize here the automatic emergence of a quantum potential Q in e.g. 
classical WDW type equations upon inserting a (Bohmian) complex wave function $\psi=
Rexp(iS/\hbar)$.  The interpretation of Q in terms of momentum fluctuations via the Fisher
information and entropy ideas is discussed along with the essentially forced role of $R^2$
as a probability density.  We also review the constructions of Padmanabhan connecting entropy
and the Einstein equations.
\end{abstract}

\maketitle

\tableofcontents

\section{ENTROPY}
\renewcommand{\theequation}{1.\arabic{equation}}
\setcounter{equation}{0}

In \cite{p1} one takes an entropy functional ($u^a=\bar{x}^a-x^a$ is a perturbation)
\bq\label{1.1} 
S=\frac{1}{8\pi}\int d^4x\sqrt{g}\left[M^{abcd}\na_au_b\na_cu_d+N_{ab}u^au^b\right]
\end{equation}
Extremizing with respect to $u_b$ leads to ($N_{ab}u^au^b=N^{ab}u_au_b$)
\bq\label{1.2}
\na_a\left(M^{abcd}\na_c\right)u_d=N^{bd}u_d
\end{equation}
Note $\int d^4x\sqrt{-g}f\na_au_b=-\int d^4x\sqrt{-g}u_b\na_af$ since via \cite{a1}
one can write
$\gd\sqrt{-g}=-(1/2)\sqrt{-g}g_{\mu\nu}\gd g^{\mu\nu}$ and $\na_ag^{\mu\nu}=0$.
Choosing M and N such that \eqref{1.2} (for all $u_d$) implies the Einstein equations
entails 
\bq\label{1.3}
M^{abcd}=g^{ad}g^{bc}-g^{ab}g^{cd};\,\,N_{ab}= 8\pi\left(T_{ab}-\frac{1}{2}g_{ab}T\right)
\end{equation}
Consequently S becomes
\bq\label{1.4}
S=\frac{1}{8\pi}\int d^4x\sqrt{-g}\left[(\na_au^b)(\na_bu^a)-(\na_bu^b)^2+N_{ab}u^au^b\right]=
\end{equation}
$$=\frac{1}{8\pi}\int
d^4x\sqrt{-g}\left[Tr(J^2)-(Tr(J))^2+8\pi\left(T_{ab}-\frac{1}{2}g_{ab}T\right)u^au^b\right]$$
where $J_a^b=\na_au^b$.  Note here
\bq\label{1.5}
\int d^4x\sqrt{-g}g^{ad}g^{bc}\na_au_b\na_cu_d=\int d^4x\sqrt{-g}(\na^au^b)(\na^cu^d)
\end{equation}
and also
\bq\label{1.6}
\na_a\left(M^{abcd}\na_c\right)u_d=\na_a\left[g^{ad}g^{bc}-g^{ab}g^{cd}\right]\na_cu_d=
\end{equation}
$$=\na_ag^{ad}g^{bc}\na_cu_d-\na_ag^{ab}g^{cd}\na_cu_d=\na_a\na^bu^a-\na^b\na_cu^c\sim
(\na_a\na^b-\na^b\na_a)u^a$$
Further (as in \eqref{1.6})
\bq\label{1.7}
M^{abcd}\na_au_b\na_cu_d=g^{ad}g^{bc}\na_au_b\na_cu_d-g^{ab}g^{cd}
\na_au_b\na_cu_d=
\end{equation}
$$=\na^du_b\na^bu_d-\na_au^a\na_cu^c$$
which confirms \eqref{1.4}.  We record also from \cite{o1} that
\bq\label{1.8}
(\na_{\mu}\na_{\nu}-\na_{\nu}\na_{\mu})\ga(w)=R(\ga,\pp_{\mu},\pp_{\nu},w)
\end{equation}
which identifies $\na_{\mu}\na_{\nu}-\na_{\nu}\na_{\mu}$ with $R_{\mu\nu}$ and allows us to imagine
\eqref{1.6} as $R_a^bu^a$ with Einstein equations
\bq\label{1.9}
R_a^bu^a=N_a^bu^a\,\,(=N^{bc}g_{ca}g^{ca}u_c)
\end{equation}
for example, which is of course equivalent to $R_{ab}=N_{ab}$ (cf. also \cite{p2}).
Note also $G_{ab}=R_{ab}-(1/2)Rg_{ab}=kT_{ab}$ implies that $R_{\nu}^{\mu}-(1/2)R\gd^{\mu}_{\nu}
=kT_{\nu}^{\mu}$ which upon contraction gives $R=-kT$ (since $\gd^{\mu}_{\mu}=4$) and hence
$R_{ab}=k(T_{ab}-(1/2)Tg_{ab})$.
\\[3mm]\indent
{\bf REMARK 1.1.}
We will insert more information on entropy and fluctuations related to equation (1.1) later based
on \cite{b3,c1,f2,h23,h24,h25,h26,h27,p3,s23} (see also Section 3 on exact uncertainty).
$\hfill\bs$
\\[3mm]\indent
For completeness we sketch here a derivation of the Einstein equations from an action principle
(cf. \cite{a1,c3,m1,w1}).  The Einstein-Hilbert action is $A=\int_{\gO}[{\mf
L}_G+ {\mf L}_M]d^4x$ where ${\mf L}_G=(1/2\chi)\sqrt{-g}{}^4R$ ($\chi=8\pi$ and ${}^4R$ is the
Ricci scalar).  Following \cite{c3} we list a few useful facts first (generally we will
write if necessary $g_{ab}T^{cb}=T^c_{\cdot a}$ and $g_{ab}T^{bc}=T^{\cdot c}_a$).
\begin{enumerate}
\item
$\na_{\gag}g^{\ga\gb}=0$ (by definitions of covariant derivative and Christoffel symbols).
\item
$\gd\sqrt{-g}=(1/2)\sqrt{-g}g^{\ga\gb}\gd g_{\ga\gb}$ and $(\gd g_{\ga\gb})g^{\ga\gb}
=-(\gd g^{\ga\gb})g_{\ga\gb}$ (see e.g. \cite{w1} for the calculation).
\item
For a vector field $v^a$ one has $\na_av^a=\pp_a(\sqrt{-g}v^a)(1/\sqrt{-g})$ and
$\na_{\gb}T^{\ga\gb}=\pp_{\gb}(\sqrt{-g}T^{\ga\gb})(1/\sqrt{-g})+\gG^{\ga}_{\gs\gb}T^{\gs\gb}$
(from $\gG^{\gs}_{\gs\ga}=(1/2)(\pp_{\ga}g_{\mu\nu})g^{\mu\nu}$ and
$\pp_{\ga}(log(\sqrt{-g})=\gG^{\gs}_{\gs\ga}$).
\item
For two metrics $g,\,g^*$ one shows that $\gd\gG^{\ga}_{\gb\gag}=\gG^{*\ga}_{\gb\gag}
-\gG^{\ga}_{\gb\gag}$ is a tensor.
\item
$\gd R_{\ga\gb}=\na_{\gs}(\gd \gG^{\gs}_{\ga\gb}-\na_{\gb}(\gd\gG^{\gs}_{\ga\gs})$
(see \cite{c3} for the calculations).
\item
Recall also Stokes theorem
$\int_{\gO}\na_{\gs}v^{\gs}\sqrt{-g}d^4x=\int_{\gO}\pp_{\gs}(v^{\gs}\sqrt{-g})d^4x=\int_{\pp\gO}
\sqrt{-g}v^{\gs}d^3\gS_{\gs}$.
\end{enumerate}
\indent
Now requiring a stationary action for arbitrary $\gd g^{ab}$ (with certain derivatives of the
$g^{ab}$ fixed on the boundary of $\gO$ one obtains  (${\mf L}_M$ is the matter Lagrangian)
\bq\label{1.10}
\gd
I=\frac{1}{2\chi}\int_{\gO}\left(R_{\ga\gb}-\frac{1}{2}g_{\ga\gb}R\right)\sqrt{-g}\gd
g^{\ga\gb}d^4x+
\end{equation}
$$+\frac{1}{2\chi}\int_{\gO}g^{\ga\gb}\sqrt{-g}\gd R_{\ga\gb}d^4x+\int_{\gO}\frac{\gd{\mf L}_M}
{\gd g^{\ga\gb}}\gd g^{\ga\gb}d^4x=0$$
The second term can be written 
\bq\label{1.11}
\frac{1}{2\chi}\int_{\gO}g^{\ga\gb}\sqrt{-g}\gd R_{\ga\gb}d^4x=\frac{1}{2\chi}\int
g^{\ga\gb}\sqrt{-g}[\na_{\gs}(\gd\gG^{\gs}_{\ga\gb})-\na_{\gb}(\gd\gG^{\gs}_{\ga\gs}]d^4x=
\end{equation}
$$=\frac{1}{2\chi}\int_{\gO}\sqrt{-g}[\na_{\gs}(g^{\ga\gb}\gd\gG^{\gs}_{\ga\gb})-\na_{\gb}
(g^{\ga\gb}\gd\gG^{\gs}_{\ga\gs})]d^4x=$$
$$=\frac{1}{2\chi}\int_{\gO}\pp_{\gs}[(\sqrt{-g}g^{\ga\gb}\gd\gG^{\gs}_{\ga\gb})-(\sqrt{-g}
g^{\ga\gs}\gd\gG^{\rho}_{\ga\rho})]d^4x$$
where $\gd\gG^{\ga}_{\gb\gag}=(1/2)[\na_{\gag}(\gd g_{\gb\gs})+\na_{\gb}(\gd g_{\gs\gag})-
\na_{\gs}(\gd g_{\gag\gb})]$.  This can be transformed into an integral over the boundary
$\pp\gO$ where it vanishes if ceertain derivatives of $g_{\ga\gb}$ are fixed on the boundary.
In fact the integral over the boundary $\pp\gO=\sum S_i$ can be written as $\sum_i(\gep_I/2\chi)
\int_{S_i}\gag_{\ga\gb}\gd \tl{N}^{\ga\gb}d^3x$ where $\gep_i={\bf n}_i\cdot{\bf n}_i=\pm 1$ 
(${\bf n}_i$ normal to $S_i$) and $\gag_{\ga\gb}=g_{\ga\gb}-\gep_i{\bf n}_{\ga}\cdot{\bf
n}_{\gb}$ is the 3-metric on the hypersurface $S_i$ (cf. \cite{y4}).  Further 
\bq\label{1.12}
\tl{N}^{\ga\gb}=\sqrt{|\gag|}(K\gag^{\ga\gb}-K^{\ga\gb})=-\frac{1}{2}g\gag^{\ga\mu}\gag^{\gb\nu}
{\mc L}_{{\bf n}}(g^{-1}\gag_{\mu\nu})
\end{equation}
where $K_{\ga\gb}=-(1/2){\mc L}_{{\bf n}}\gag_{\ga\gb}$ is the extrinsic curvature of each
$S_i$ and ${\mc L}_{{\bf n}}$ is the Lie derivative.  Consequently if the quantities 
$\tl{N}^{\ga\gb}$ are fixed on the boundary for an arbitrary $\gd g_{\ga\gb}$ one gets from
the first and last equations in \eqref{1.10} the Einstein field equations
\bq\label{1.13}
G_{\ga\gb}=R_{\ga\gb}-\frac{1}{2}Rg_{\ga\gb}=\chi T_{\ga\gb};\,\,T_{\ga\gb}=-2\frac{\gd{\mf
L}_M}{\gd g^{ab}}+{\mf L}_Mg_{\ga\gb}
\end{equation}
We note here that 
\bq\label{1.14}
\gd\int {\mf L}_m\sqrt{-g}d^4x=\int\frac{\gd{\mf L}_m}{\gd g^{ab}}\sqrt{-g}d^4x+\int{\mf
L}_m\gd(\sqrt{-g})d^4x=
\end{equation}
$$=\int\frac{\gd{\mf L}_m}{\gd g^{ab}}\sqrt{-g}d^4x-\frac{1}{2}\int{\mf L}_mg_{ab}
(\gd g^{ab})\sqrt{-g}d^4x$$
A factor of 2 then arises from the $2\chi$ in (1.10).
\\[3mm]\indent
{\bf REMARK 1.2.}
Let us rephrase some of this following \cite{w1} for clarity.  Thus e.g. think of functionals
$F(\psi)$ with $\psi=\psi_{\gl}$ a one parameter family and set $\gd\psi=(d\psi_{\gl}/d\gl)|_
{\gl=0}$.  For $F(\psi)$ one writes then $dF/d\gl=\int \phi\gd\psi$ and sets $\phi=
(\gd F/\gd\psi)|_{\psi_0}$.  Then (assuming all functional derivatives are symmetric with
no loss of generality) one has for ${\mf L}_G=\sqrt{-g}R$ and $S_G=\int {\mf L}_Gd^4x$
\bq\label{1.15}
\frac{d{\mf L}_G}{d\gl}=\sqrt{-g}(\gd R_{ab})g^{ab}+\sqrt{-g}R_{ab}\gd g^{ab}+R\gd(\sqrt{-g})
\end{equation} 
But $g^{ab}\gd R_{ab}=\na^av_a$ for $v_a=\na^b(\gd g_{ab})-g^{cd}\na_a(\gd g_{cd})$.  Further
$\gd\sqrt{-g}=-(1/2)\sqrt{-g}g_{ab}\gd g^{ab}$ so one has
\bq\label{1.16}
\frac{dS_G}{d\gl}=\int\frac{d{\mf L}_G}{d\gl}d^4x=\int\na^av_a\sqrt{-g}d^4x+\int
\left(R_{ab}-\frac{1}{2}Rg_{ab}\right)(\gd g^{ab})\sqrt{-g}d^4x
\end{equation}
Discarding the first term as a boundary integral we get the first term in (1.10).$\hfill\bs$
\\[3mm]\indent
{\bf REMARK 1.3.}
From \cite{c7,p1} we see that the entropy in S in (1.1) reduces to a 4-divergence when the 
Einstein equations are satisfied ``on shell" making S a surface term
\bq\label{2.77}
S=\frac{1}{8\pi}\int_Vd^4x\sqrt{-g}\na_i(u^b\na_bu^i-u^i\na_bu^b)=
\end{equation}
$$=\frac{1}{8\pi}
\int_{\pp V}d^3x\sqrt{h}n_i(v^b\na_bu^i-u^i\na_bu^b)$$
Thus the entropy of a bulk region V of spacetime resides in its boundary $\pp V$
when the Einstein equations are satisfied.  In varying (1.1) to obtain
(1.2) one keeps the surface contribution to be a constant.  Thus in a
semiclassical limit when the Einstein equations hold to the lowest order
the entropy is contributed only by the boundary term and the system is holographic.
$\hfill\bs$

\section{WDW}
\renewcommand{\theequation}{2.\arabic{equation}}
\setcounter{equation}{0}

We gather now some information about the derivation of Einstein's equations from an action principle
and also discuss the Hamiltonian theory involving the Einstein-Hamilton-Jacobi (EHJ) equation and the
WDW equation.  One recalls from \cite{g1} that the EHJ equation
\bq\label{2.1}
{}^3R+\frac{1}{h}\left(\frac{1}{2}h_{ij}h_{k\ell}-h_{ik}h_{j\ell}\right)\left(
\frac{\gd S}{\gd h_{ij}}\right)\left(\frac{\gd S}{\gd h_{k\ell}}\right)=0
\end{equation}
($h_{ij}$ corresponds to the metric of a spatial hypersurface) plus a principle of constructive
interference of deBroglie waves leads to the entire set of 10 Einstein equations.  The idea of
Tomonaga's multi fingered time is used here (cf. also \cite{c1,n1}).
\\[3mm]\indent
Now there are a number of derivations of the WDW equations with connections to Bohmain dynamics
and the quantum potential in \cite{c1,h4,hal,pnt,pss,s7,s9,s13,s14,s16,s21} and we will go 
directly to \cite{h4,hal} after a few comments.
First let us recall the deWitt metric for which we refer to
\cite{b1,c1,d1,f1,g2,g3,g4,h4,hal,h1,k1,k2,k3,m1,pnt,pss,r1,s7,s9,s13,s14,s16,s21,w2}
(cf. also \cite{s50,w3}).  Various
formulas arise for the WDW which involve a deWitt metric (or supermetric) 
\bq\label{2.2}
G_{abcd}^{\ga}=\frac{1}{\sqrt{h}}(h_{ac}h_{bd}+h_{ad}h_{bc}-2\ga h_{ab}h_{cd});
\end{equation}
$$G^{abcd}_{\gb}=\frac
{\sqrt{h}}{2}(h^{ac}h^{bd}+h^{ad}h^{bc}-2\gb h^{ab}h^{cd})$$
where $\ga+\gb=3\ga\gb$.  For general relativity (GR) one takes $\gb=1$ and $\ga=1/2$ (see e.g.
\cite{g2,g3,k1} for this form of metric).  Here the WDW equaiton for GR is ($c=1$)
\bq\label{2.3}
{\mf H}\psi[h_{ab},\phi]=[-16\pi G\hbar^2G_{abcd}\frac{\gd^2}{\gd h_{ab}\gd h_{cd}}-
\frac{\sqrt{h}}{16\pi G}(R-2\gL)+{\mf H}_m]\psi=0
\end{equation}
where $h_{ab}$ is a 3-metric, R the 3-D Ricci scalar, $\gL$ the cosmological constant,
$G_{abcd}=G^{1/2}_{abcd}$ the deWitt metric, and ${\mf H}_m$ is the Hamiltonian density for
non-gravitational fields.  The integrated form of \eqref{2.3} is
\bq\label{2.4}
\int d^3x\,N{\mf H}\psi={\mf H}^N\psi=\left({\mf H}^N_G+{\mf H}^N_m\right)\psi=0
\end{equation}
Writing $\psi=exp\left(\frac{i(MS_0+S_1+M^{-1}S_2+\cdots}{\hbar}\right)$ for $M=(32\pi G)^{-1}$ leads to
a power series in M with second term
\bq\label{2.5}
{\mf H}_x=\frac{1}{2}G_{abcd}\frac{\gd S_0}{\gd h_{ab}}\frac{\gd S_0}{\gd h_{cd}}-2\sqrt{h}(R-2\gL)=0
\end{equation}
which is the Hamilton-Jacobi (HJ) equation for the gravitational field and we refer to \cite{g2,g3,k1} for
more details.
\\[3mm]\indent
It will be important to see here how the quantum potential arises and we go to \cite{pnt,pss} with a metric
\bq\label{2.6}
ds^2=-(N^2-N^iN_i)dt^2+2N_idx^idt+h_{ij}dx^idx^j
\end{equation}
(classical ADM situation - cf. \cite{a2,b2,m1}) and Hamiltonian
\bq\label{2.7}
H=\int d^3x (N{\mf H}+N^j{\mf H}_j);\,\,{\mf H}_j=-2D_i\pi_j^i+\pi_{\phi}\pp_j\phi;
\end{equation}
$${\mf H}=\gk
G_{ijk\ell}\pi^{ij}\pi^{k\ell}+\frac{1}{2}h^{-1/2}\pi_{\phi}^2+h^{1/2}\left[-\gk^{-1}({}^3R-2\gL)
+\frac{1}{2}h^{ij}\pp_i\phi\pp_j\phi+U(\phi)\right]$$
where $\gk=16\pi G/c^4$, $D_k$ is the covariant derivative, and
\bq\label{2.8}
\pi^{ij}=-h^{1/2}\left(K^{ij}-h^{ij}K\right)=G^{ijk\ell}[\dot{h}_{k\ell}-D_kN_{\ell}-D_{\ell}N_k];
\end{equation}
$$K_{ij}=-\frac{1}{2N}(\dot{h}_{ij}-D_iN_j-D_jN_i)$$
Thus $K_{ij}$ is the extrinsic curvature of the hypersurface and
$(\clubsuit)\,\,\pi_{\phi}=(h^{1/2}/N)(\dot{\phi}-N^i\pp_j\phi)$ where $\phi$ is a matter field.
The classical 4-metric above and the scalar field which are solutions of the Einstein equations
can be obtained from the Hamiltonian equations of motion
\bq\label{2.9}
\dot{h}_{ij}=\{h_{ij},H\};\,\,\dot{\pi}^{ij}=\{\pi^{ij},H\};\,\,\dot{\phi}=\{\phi,H\};\,\,
\dot{\pi}_{\phi}=\{\pi_{\phi},H\}
\end{equation}
for some choice of N and $N^j$, given suitable initial conditions compatible with the constraints
$(\spadesuit)\,\,{\mf H}\approx 0$ and ${\mf H}_j\approx 0$ (in standard terminology).
There is a standard constraint algebra involving Poisson brackets of the ${\mf H}_i$ and ${\mf H}$
(see e.g. \cite{pnt}) and for quantization the constraints become conditions on the possible states of the
quantum system yielding equations $(\bullet)\,\,\hat{{\mf H}}_i|\psi>=0$ and $\hat{{\mf H}}|\psi>=0$
leading to $(\bl)\,\,-2h_{ij}D_j[\gd\psi/\gd h_{ij}]+[\gd\psi/\gd\phi]\pp_i\phi=0$ and the WDW equation
\bq\label{2.10}
\left\{-\hbar^2\left[\gk G_{ijk\ell}\frac{\gd}{\gd h_{ij}}\frac{\gd}{\gd
h_{k\ell}}+\frac{1}{2}h^{-1/2}\frac{\gd^2}{\gd\phi^2}\right]+V\right\}\psi(h_{ij},\phi)=0;
\end{equation}
$$V=h^{1/2}\left[-\gk^{-1}({}^3R-2\gL)+\frac{1}{2}h^{ij}\pp_i\pp_j\phi+U(\phi)\right]$$
This involves products of local operators at the same space point so regularization is indicated
(we omit details).  
\\[3mm]\indent
Now for the Bohmian point of view one writes $\psi=Aexp(iS/\hbar)$ where $A$ and $S$ are functionals of 
$h_{ij}$ and $\phi$ leading to two equations indicating that A and S are invariant under general space
coordinate transformations, namely
\bq\label{2.11}
-2h_{ij}D_j\frac{\gd S}{\gd h_{ij}}+\frac{\gd S}{\gd\phi}\pp_i\phi=0;\,\,-2h_{ij}D_j\frac{\gd A}{\gd h_{ij}}
+\frac{\gd A}{\gd\phi}\pp_i\phi=0
\end{equation}
These could depend on factor ordering but in any event one will have e.g. the form
\bq\label{2.12}
\gk G_{ijk\ell}\frac{\gd S}{\gd h_{ij}}\frac{\gd S}{\gd h_{k\ell}}+\frac{1}{2}h^{-1/2}\left(\frac{\gd
S}{\gd \phi}\right)^2+V+Q=0;
\end{equation}
$$Q=-\frac{\hbar^2}{A}\left(\gk G_{ijk\ell}\frac{\gd^2A}{\gd h_{ij}\gd h_{k\ell}}+\frac{h^{-1/2}}{2}\frac
{\gd ^2A}{\gd\phi^2}\right)$$
where the unregularied Q above depends on the regularization and factor ordering prescribed for the WDW
equation.  In addition to \eqref{2.12} one has
\bq\label{2.13}
\gk G_{ijk\ell}\frac{\gd}{\gd h_{ij}}\left(A^2\frac{\gd S}{\gd h_{k\ell}}\right)+\frac{h^{-1/2}}{2}
\frac{\gd}{\gd\phi}\left(A^2\frac{\gd S}{\gd\phi}\right)=0
\end{equation}
One can stipulate that the 3-metric of spacelike hypersurfaces, the scalar field, and their canonical
momenta always exist and the metric and scalar field can be determined via guidance relations
\bq\label{2.14}
\pi^{ij}=\frac{\gd S}{\gd h_{ij}};\,\,\pi_{\phi}=\frac{\gd S}{\gd\phi}
\end{equation}
with $\pi^{ij}$ and $\pi_{\phi}$ given via \eqref{2.8} etc.  Note that one cannot interpret \eqref{2.13}
as a continuity equation for a probability density due to the hyperbolic nature of the deWitt metric.
Note also that whatever may be the form of Q it must be a scalar density of weight one; indeed
from \eqref{2.12}
\bq\label{2.15}
Q=-\gk G_{ijk\ell}\frac{\gd S}{\gd h_{ij}}\frac{\gd S}{\gd h_{k\ell}}-\frac{h^{-1/2}}{2}\left(
\frac{\gd S}{\gd \phi}\right)^2-V
\end{equation}
and we refer to \cite{pnt} for the arguments.  In addition note that Q can depend only on $h_{ij}$ and 
$\phi$.

\section{EXACT UNCERTAINTY}
\renewcommand{\theequation}{3.\arabic{equation}}
\setcounter{equation}{0}

We go now to \cite{c1,h4,hal,r5,r6} and show how the WDW equation can be derived from a so
called exact uncertainty principle of Hall and Reginatto.  The idea here is that uncertainty
can be promoted to be the fundamental element distinguishing quantum and classical mechanics. 
In this approach  nonclassical fluctuations are added to the deterministic connection between
position and momentum (via the uncertainty principle) one essentially generates the quantum
potential.  In \cite{h4,hal} this is applied to gravity and a WDW equation is derived and
originally this approach was used to generate the Schr\"odinger equation (SE) (see Remark
5.2 for additional clarifications following \cite{r5,r6}). 
Thus take a
metric as in \eqref{2.6} and think of the metric $h_{ij}$ as being imprecise with a
probability distribution $P[h_{ij})$.  Take a single field classical Hamiltonian of the form
\bq\label{3.1}
H_0[h_{ij},\pi^{ij}]=\int dx\left[N\left(\frac{1}{2}G_{ijk\ell}\pi^{ij}\pi^{k\ell}+V(h_{ij})\right)-
2N_i\na_j\pi^{ij}\right]
\end{equation}
(here $D_j\sim\na_j$ is the covariant derivative).  As an ensemble Hamiltonian one takes now
\bq\label{3.2}
\tl{H}_c[P,S]=\int Dh\,PH_0[h_{ij},(\gd S/\gd h_{ij})]
\end{equation}
leading to equations of motion
\bq\label{3.3}
\pp_tP+\int dx\frac{\gd}{\gd h_{ij}}(P\dot{h}_{ij})=0;\,\,\pp_tS+H_0[h_{ij},(\gd S/\gd h_{ij})]=0;
\end{equation}
$$\dot{h}_{ij}=NG_{ijk\ell}\frac{\gd S}{\gd h_{k\ell}}-\na_jN^i-\na_iN_j$$
The lack of conjugate momenta for the lapse and shift components N and $N_i$ places constraints on the
classical equations of motion which in the ensemble formalism take the form
\bq\label{3.4}
\frac{\gd P}{\gd N}=\frac{\gd P}{\gd N_i}=\frac{\pp P}{\pp t}=0;\,\,\na_j\left(\frac{\gd P}{\gd
h_{ij}}\right)=0;
\end{equation}
$$\frac{\gd S}{\gd N}=\frac{\gd S}{\gd N_i}=\frac{\pp S}{\pp t}=0;\,\,\na_j\left(\frac{\gd S}{\gd
h_{ij}}\right)=0$$
This corresponds to invariance of the dynamics with respect to N, $N_i$, and the initial time; also
to invariance of P and S under arbitrary spatial coordinate transformations.  Applying these constraints
to the above classical equations for the ``Gaussian" choice $N=1$ and $N_i=0$ yields
\bq\label{3.5}
\frac{\gd}{\gd h_{ij}}\left(PG_{ijk\ell}\frac{\gd S}{\gd h_{k\ell}}\right)=0;\,\,\frac{1}{2}G_{ijk\ell}
\frac{\gd S}{\gd h_{ij}}\frac{\gd S}{\gd h_{k\ell}}+V=0;\,\,V\sim c\sqrt{h}(2\gL-{}^3R)
\end{equation}
Now the exact uncertainty approach involves writing $(\bgs)\,\,\pi^{ij}=(\gd S/\gd h_{ij})+f^{ij}$ where
$f^{ij}$ vanishes on average for all configurations.  This adds a kinetic term to the average ensemble
energy leading to
\bq\label{3.6}
\tl{H}_q=<E>=\tl{H}_c+\frac{1}{2}\int Dh\,P\int dx\,NG_{ijk\ell}\overline{f^{ij}f^{k\ell}}
\end{equation}
Note here that the term in \eqref{3.1} which is linear in the derivative of $\pi^{ij}$ can be integrated
by parts giving a term directly proportional to $\pi^{ij}$ which remains unchanged when the fluctuations
are added and averaged.  Now using some general properties of causality, independence, invariance, and
exact uncertainty (cf. \cite{c1,h4,hal}) one arrives at
\bq\label{3.7}
\tl{H}_q[P,S]=\tl{H}_c[P,S]+\frac{c}{2}\int Dh\int dx\, NG_{ijk\ell}\frac{1}{P}\frac{\gd P}{\gd
h_{ij}}\frac{\gd P}{\gd h_{k\ell}}
\end{equation}
where $C$ is a positive universal constant.  Now if one defines $\hbar=2\sqrt{c}$ and
$\psi[h_{ij}]=\sqrt{P}exp(iS/\hbar)$ then,
calculating as above, we obtain a WDW equation for quantum geometry in the form
\bq\label{3.8}
\left[-\frac{\hbar^2}{2}\frac{\gd}{\gd h_{ij}}G_{ijk\ell}\frac{\gd}{\gd
h_{k\ell}}+V\right]\psi=0
\end{equation}
with a Q term $-(\hbar^2/2P)G_{ijk\ell}(\gd^2P/\gd h_{ij}\gd h_{k\ell})$ added in the
Hamiltonian equation (cf. (2.12)).  Note further
\bq\label{3.9}
\frac{\gd\psi}{\gd N}=\frac{\gd \psi}{\gd N_i}=\frac{\pp\psi}{\pp t}=0;\,\,\na_j\left(\frac{\gd\psi}{\gd
h_{ij}}\right)=0
\end{equation}
An important feature of this WDW equation is that it is obtained with a particular operator ordering.
Indeed $G_{ijk\ell}$ is sandwiched between the two functional derivatives and thus ambiguity is removed
in this respect.  One recalls that the same thing happens with the SE which is derived in the form
$(\bullet\bullet)\,\,i\hbar\pp_t\psi=-(\hbar^2/2)\na\cdot(1/m)\na\psi+V\psi$.
\\[3mm]\indent
Now in the theory of Schr\"odinger equations there is a strong connection between terms of the
form
\bq\label{3.10}
I=\frac{1}{2}g^{ik}\int \frac{1}{P}\frac{\pp P}{\pp y^i}\frac{\pp P}{\pp y^k}d^ny
\end{equation}
and concepts of Fisher information, entropy, and quantum potential (see \cite{c1} for an extensive
development).  Classical Fisher information is known to be connected to various forms of entropy via
formulas like (cf. \cite{c1,g8})
\bq\label{3.11}
\frac{\pp{\mf S}}{\pp t}=\frac{\hbar}{2m}{\mf F}=\frac{\hbar}{2m}\int\frac{(\na \rho)^2}{\rho}=
\frac{4}{\hbar}\int \rho Q;\,\,Q=-\frac{\hbar^2}{2m}\frac{\gd\sqrt{\rho}}{\sqrt{\rho}}
\end{equation}
Here ${\mf S}\sim -\int \rho log(\rho)$ is a so-called differential entropy and $\rho$ here corresponds to 
P or A in the notation of this paper (note P and A refer to 3-space quantities); ${\mf F}$ is a
Fisher information measure.  There are relations between differential entropy and
Shannon-Boltzman entropy for example and we refer to
\cite{c1,c7,c8,g1} for details.  We remark also that Olavo in \cite{o3} derives Schr\"odinger
equations using entropy ideas where the entropy in \cite{o3} is of Shannon-Boltzman type ${\mc
S}=k_Blog(W)=-k_Blog(P)$ where $P=1/W$ is the probability of a microstate occurance.  One deals
with momentum fluctuations $\overline{(\gd p)^2}$ and assumes $\overline{(\gd p)^2(\gd
x)^2}=\hbar^2/4$.  There results $\overline{(\gd p)^2}\sim -(\hbar^2/4)\pp^2log(\rho)$ where
$\rho$ is a probability density.  Then ${\mc S}_{equilib}\sim k_Blog(\rho)$ implies
$\overline{(\gd p)^2}\sim -(\hbar^2/4k_B)\pp^2{\mc S}_{equilib}$.  Note also here that
(calculating in 1-D for convenience) $\pp^2log(\rho)=(\rho''/\rho)-(\rho'/\rho)^2$ and 
\bq\label{3.12}
Q=-\frac{\hbar^2}{2m}\frac{\pp^2\sqrt{\rho}}{\sqrt{\rho}}=\frac{\hbar^2}{8m}\left[2\frac{\rho''}{\rho}
-\left(\frac{\rho'}{\rho}\right)^2\right]=
\frac{\hbar^2}{8m}\left[2\pp^2{\mc S}_{equilib}+(\pp{\mc S}_{equilib})^2\right]
\end{equation}
The theme here is to relate entropy, the quantum potential, and geometry in the relativistic context.
One can think of entropy or of quantum fluctuations as generating quantum behavior (often via a quantum
potential) and we want to connect these matters to the Einstein equations in a Bohmian spirit.  Most of
this is already done and sketched in \cite{c1} for example and we want 
to make it more explicit here (see Section 5.1 for some clarifications of the exact
uncertainty method and WDW).
\\[3mm]\indent
{\bf REMARK 3.1.}
We call attentiion here to \cite{c30,c31} (cf. also \cite{l1,m7,y5}) where Fisher information
is related to uncertainty relations and a differential Shannon entropy is introduced.
$\hfill\bs$

\section{WDW AND THE QUANTUM POTENTIAL}
\renewcommand{\theequation}{4.\arabic{equation}}
\setcounter{equation}{0}

We have seen already how the quantum potential arises in the WDW equation in Section 2 following 
\cite{pnt,pss}.  Let us now approach this from another point of view following \cite{s14,s16,s21}
(see also \cite{c1}).  One takes $16\pi G=1$ here for convenience and uses the standard ADM decomposition.
Assuming first that there is no matter field the Lagrangian density of GR is
\bq\label{4.1}
{\mf L}=\sqrt{-g}R=\sqrt{h}N[{}^3R+Tr(K^2)-(tr(K))^2]
\end{equation}
($K_{ij}$ is the extrinsic curvature).  The canonical momentum of the 3-metric is 
\bq\label{4.2}
p^{ij}=\frac{\pp{\mf L}}{\pp\dot{h}_{ij}}=\sqrt{h}[K^{ij}-h^{ij}Tr(K)]
\end{equation}
The classical Hamiltonian is $H=\int d^3x{\mf H}$ where ${\mf H}=\sqrt{h}(NC+N^iC_i)$ and the constraints
are
\bq\label{4.3}
C=-{}^3R+\frac{1}{h}\left[Tr(p^2)-\frac{1}{2}(Tr(p))^2\right]=-2G_{\mu\nu}n^{\mu}n^{\nu};
\end{equation}
$$C_i=-2{}^3\na^j\left(\frac{p_{ij}}{\sqrt{h}}\right)=-2G_{\mu i}n^{\mu}$$
where $n^{\mu}$ is the normal to the spatial hypersurfaces $n^{\mu}=(1/N,-(1/N){\bf N})$.
As usual in a Bohmian theory one adds a quantum potential to the Hamiltonian in order to obtain the correct
equations of motion so $H\to H+Q$ or ${\mf H}\to {\mf H}+{\mf Q}$ where $Q=\int d^3x{\mf Q}$ and the
quantum potential is
\bq\label{4.4}
{\mf Q}=\hbar^2NG_{ijk\ell}\frac{1}{|\psi|}\frac{\gd^2|\psi|}{\gd h_{ij}\gd h_{k\ell}}
\end{equation}
This means that one must modify the classical constraints via $(\bgs\bgs)\,\,C\to C+({\mf
Q}/\sqrt{h}N)$ and
$C_i\to C_i$.  For the constraint algebra one considers
\bq\label{4.5}
C(N)=\int d^3x\sqrt{h}NC;\,\,\tl{C}({\bf N})=\int d^3x\sqrt{h}N^iC_i
\end{equation}
and there results (cf. \cite{c1,s14,s16})
\bq\label{4.6}
\{\tl{C}({\bf N}),\tl{C}({\bf N}')\}=\tl{C}({\bf N}\cdot\na{\bf N}'-{\bf N}'\cdot\na{\bf N});
\end{equation}
$$\{\tl{C}({\bf N}),C(N)\}=C({\bf N}\cdot\na N);\,\,\{C(N),C(N')\}\sim 0$$
The first 3-diffeomorphism subalgebra has no change relative to the classical equation and the second,
representing the fact that the Hamiltonian constraint is a scalar under 3-diffeomorphisms, is also the same
as the classical situation.  In the third case the quantum potential changes the Hamiltonian constraint
algebra dramatically (making it weakly equal to zero).  The details are written out in \cite{s14,s16}
using the Bohm-Hamilton-Jacobi equation
\bq\label{4.8}
G_{ijk\ell}p^{ij}p^{k\ell}+\sqrt{h}({}^3R-2\gL)+{\mf Q}=0
\end{equation}
which is differentiated to obtain
\bq\label{4.9}
\frac{1}{N}\frac{\gd}{\gd h_{ij}}\frac{{\mf
Q}}{\sqrt{h}}=\frac{3}{4\sqrt{h}}h_{k\ell}p^{ij}p^{k\ell}\gd(x-z)-\frac{\sqrt{h}}{2}h^{ij}({}^3R-2\gL)
\gd(x-z)-\sqrt{h}\frac{\gd {}^3R}{\gd h_{ij}}
\end{equation}
Putting this information in the Poisson brackets one obtains the last relation in \eqref{4.6}.  
The existence of the quantum potential shows that the quantum algebra is a 3-diffeomorphism algebra times
an Abelian subalgebra and the only difference with \cite{m5} for example is that this algebra is weakly
closed (this will eventually mean closed on the Bohmian trajectories).
Thus the algebra \eqref{4.6} is a projection of general coordinate transformations
to the spatial and temporal diffeomorphisms and the equations of motion are invariant under such
transformations.  The important point here is that, although the form of the quantum potential depends on
the regularization and ordering, nevertheless in the quantum constraints algebra the form of the quantum
potential is not important; the equations are correct independently of the definition of quantum potential.
Note that for $16\pi G=1$ and $c=1$ (so $\gk=1$) we can relate (4.7) and (2.12) provided
there is no matter field, $N=1$, and we assume a different signature in the metric.
\\[3mm]\indent
Now one can derive the quantum corrections to the Einstein equations.  
For the dynamical part consider
\bq\label{4.10}
\dot{h}^{ij}=\{H,h^{ij}\};\,\,\dot{p}_{ij}=\{H,p_{ij}\}
\end{equation}
and some calculation which we omit leads to
$(\bgs\bgs)\,\,{\mf G}^{ij}=-(1/N)(\gd Q/\gd h_{ij})$ which means that the quantum force modifies
the dynamical parts of the Einstein equations.  For the nondynamical parts one uses the
constraint equations \eqref{4.3} to obtain ${\mf G}^{00}={\mf Q}/2N^3\sqrt{h}$ and ${\mf
G}^{0i}=-{\mf Q}N^i/2N^3\sqrt{h}$, which can be written in the form $(\bullet\bullet)\,\,{\mf
G}^{0\mu}=({\mf Q}/2\sqrt{h})g^{0\mu}$. The equations obtained via the Hamiltonian will also
agree with those given by the phase of the wave function and the guidance formula.  Indeed from
the Bohmian HJ equation \eqref{4.8} one has 
\bq\label{4.11}
G_{ijk\ell}\frac{\gd S}{\gd h_{ij}}\frac{\gd S}{\gd h_{k\ell}}-\sqrt{h}({}^3R-{\mf Q})=0
\end{equation}
To get the equation of motion one must differentiate the HJ equation with respect to $h_{ab}$ and use the
guidance formula $p^{k\ell}=\sqrt{h}(K^{k\ell}-h^{k\ell}K)=\gd S/\gd h_{k\ell}$ and doing this leads again
to $(\bgs\bgs)$ so the evolution generated by the Hamiltonian is compatible with the guidance
formula.
\\[3mm]\indent
Inclusion of matter fields is straightforward; one simply adds the matter quantum potential and writes
\bq\label{4.12}
{\mf G}^{ij}=-\gk{\mf T}^{ij}-\frac{1}{N}\frac{\gd(Q_G+Q_m)}{\gd g_{ij}};\,\,{\mf G}^{0\mu}=-\gk{\mf
T}^{0\mu}+\frac{{\mf Q}_G+{\mf Q}_m}{2\sqrt{-g}}g^{0\mu}
\end{equation}
Here ($\phi$ is the matter field)
\bq\label{4.13}
{\mf Q}_m=\hbar^2\frac{N\sqrt{h}}{2}\frac{1}{|\psi|}\frac{\gd^2|\psi|}{\gd\phi^2};\,\,
{\mf Q}_G=\hbar^2NhG_{ijk\ell}\frac{1}{|\psi|}\frac{\gd^2|\psi|}{\gd h_{ij}\gd h_{k\ell}}
\end{equation}
and $Q_G=\int d^3x{\mf Q}_G$ with $Q_m=\int d^3x{\mf Q}_m$.  The equations \eqref{4.12}
are the Bohm-Einstein equations and are the quantum version of the Einstein equations.  Since regularization
only affects the quantum potential the quantum Einstein equations are the same for any regularization.
They are invariant under temporal $\ot$ spatial diffeomorphisms and can be written also in the form
\bq\label{4.14}
{\mf G}^{\mu\nu}=-\gk{\mf T}^{\mu\nu}+{\mf S}^{\mu\nu};\,\,{\mf S}^{\mu\nu}=\frac{{\mf
Q}}{2\sqrt{-g}}g^{0\mu};\,\,{\mf S}^{ij}=-\frac{1}{N}\frac{\gd Q}{\gd g_{ij}}
\end{equation}
We refer to \cite{c1,s14,s16,s21} for further discussion.

\section{FISHER INFORMATION AND ENTROPY}
\renewcommand{\theequation}{5.\arabic{equation}}
\setcounter{equation}{0}

We will connect up here various ideas of entropy and Fisher information 
(following \cite{c1,f17}).
First recall $N_{ab}$ in Section 1 corresponds to 
$T_{ab}-(1/2)g_{ab}T$ and one can imagine this arising from a matter Lagrangian ${\mf L}_m$
as in (1.10)-(1.13) where $G_{ab}=R_{ab}-(1/2)Rg_{ab}=\chi T_{ab}\sim \chi(-2(\gd{\mf L}/\gd
g^{ab}) +{\mf L}_mg_{ab})$.  We recall that $R=-\chi T$ so $R_{ab}=\chi(T_{ab}-(1/2)g_{ab}T)$
and note that
\bq\label{5.1}
T=g^{ab}{\mf L}_mg_{ab}-2g^{ab}\frac{\gd{\mf L}_m}{\gd g^{ab}}={\mf L}_m-2g^{ab}\frac
{\gd{\mf L}_m}{\gd g^{ab}}
\end{equation}
Hence 
\bq\label{5.2}
R_{ab}=\chi\left({\mf L}_mg_{ab}-2\frac{\gd{\mf L}_m}{\gd g^{ab}}-\frac{1}{2}\left[
{\mf L}_m-2g^{ab}\frac{\gd {\mf L}_m}{\gd g^{ab}}\right]\right)=
\end{equation}
$$=\chi\left(\frac{1}{2}g_{ab}{\mf L}_m-\frac{\gd{\mf L}_m}{\gd g^{ab}}\right)$$
and in the situation of Section 1 we have $R_{ab}u^b\sim N_{ab}u^b$.  It is clear however
from Sections 2-4 that one does not need a matter potential in order to discuss the quantum
potential in general spaces.
\\[3mm]\indent
One goes to the deWitt 6-dimensional ``superspace" with metric $G_{ijk\ell}$ (cf. \cite{d1})
(here $G_{ijk\ell}=(1/2\sqrt{h})(h_{ik}h_{j\ell}+h_{i\ell}h_{jk}-h_{ij}h_{k\ell})$ following
\cite{d1} - cf. also (2.2) which differs by a factor of 2).  The Fisher information will have a
general form
\bq\label{5.3}
I=4\int d{\bf x}\int Dg\sum_{ijk\ell}G_{ijk\ell}\frac{\pp\psi^*}{\pp
h_{ij}}\frac{\pp\psi}{\pp h_{k\ell}}
\end{equation}
where $d{\bf g}\sim \prod dg_{ij}$ (cf. \cite{c20,c1,f17,f19,p11}).  To motivate and clarify
this one thinks of a probability density function $f(y|\gt)$ used in estimating a
parameter $\gt$ based on imperfect observations $y=\gt+x$ ($x\sim$ noise).  Assume unbiased
estimates, namely $(\bl\bl)\,\,<\hat{\gt}(y)-\gt>=0=\int dy[\hat{\gt}-\gt]p(y|\gt)$ where
$p(y|\gt)$ is the probability for y in the presense of one parameter value $\gt$. 
Differentiate $(\bl\bl)$ to get $\int dy(\hat{\gt}-\gt)\pp_{\gt}p-\int dy p=0$ and via
$\int p=1$ and $\pp_{\gt}=p\pp_{\gt}log(p)$ one arrives at
\bq\label{5.4}
\int dy(\hat{\gt}-\gt)\pp_{\gt}log(p)=1=\int
dy[\sqrt{p}\pp_{\gt}log(p)][(\hat{\gt}-\gt)\sqrt{p}]
\end{equation}
The Schwartz inequality gives then
\bq\label{5.5}
\int dy(\pp_{\gt}log(p))^2p\int dy (\hat{\gt}-\gt)^2p\geq 1
\end{equation}
(Cramer-Rao inequality) which links the mean square estimate $e^2$ (second
factor) to the Fisher information I (first factor).  In \cite{f19} one writes $p=q^2$
so $(\spadesuit\spadesuit)\,\,I\sim 4\int dx(q')^2$ and various quadratic Lagrangians in
physics are considered, e.g. $(1)\,\,(1/2)m(\dot{q})^2-V,\,\,
(2)\,\,-\na\psi\cdot\na\psi^*+\cdots,\,\,
(3)\,\,-(\hbar^2/2m)\na\psi\cdot\na\psi^*+\cdots,\,\,(4)\,\,\sum
g_{mn}(q(\tau))\pp_{\tau}q_m\pp_{\tau}q_n,$ etc.  A principle of extreme physical information
(EPI) is ennunciated (in a game theoretic context) and, setting e.g.
$x_1=ix,\,\,x_2=iy,\,\,x_3=iz,$ and $x_4=ct$ with $(x_1,x_2,x_3)\sim{\bf r}$,
one posits modes $q_n=q_n({\bf r},t)$ with $\psi_n=q_{2n-1}+iq_{2n}$ ($n=1,\cdots,
N/2)$).  Then take $(\clubsuit\clubsuit)\,\,\sum_1^{N/2}\psi_n^*\psi_n=\sum q_n^2=
p({\bf r},t)$ with $I\sim 4\sum_1^{N/2}\int d\tau\na q_n\cdot\na q_n$ (cf. $(\spadesuit
\spadesuit)$).
Then physical content is introduced via Fourier transform momentum-energy variables
$(i{\bf r},ct)\leftrightarrow [(i\mu/\hbar),(E/ct)]$ with $\psi_n\leftrightarrow\phi_n$
so that $(\na\psi_n,\pp_t\psi_n)\leftrightarrow [(-i\mu\phi_n/\hbar),(iE\phi_n/\hbar)]$ (only
$E=mc^2$ will be assumed physically below).  I is regarded as information obtained by an
observer and this is to be balanced by the physical payoff J by a ``demon" expressed in
physical terms.  The net information change $\gD I=I-J$ should be zero (as in zero sum game)
and EPI specifies that $I=J$ which means here ($\sum_1^{N'/2}\phi_n^*\phi_n=P(\mu,E)$)
\bq\label{5.6}
I=4c\sum_1^{N/2}\int\int d\tau dt\left[-(\na\psi_n)^*\cdot\na\psi_n+\left(\frac{1}{c^2}\right)
\left(\frac{\pp\psi_n}{\pp t}\right)^*\left(\frac{\pp\psi_n}{\pp t}\right)\right]=I=
\end{equation}
$$=J=\frac{4c}{\hbar^2}\int\int d\mu
dE\,P(\mu,E)(-\mu^2+(E^2/c^2)=\frac{4c}{\hbar^2}\left<-\mu^2+\frac{E^2}{c^2}\right>$$
Some argument then gives $-\mu^2+(E^2/c^2)=m^2c^2$ and minimizing $\gD I$ ($\gD I=0$)
leads to the Klein-Gordon equation.  This approach seems a little silly but it is also cute;
it does in any case sort of motivate the use of (5.3) as a Fisher information.
\\[3mm]\indent
{\bf REMARK 5.1.}
The entropy in Section 1 is of course contrived via perturbations in displacement
and their derivatives (elastic deformation) and is not designed for quantization
(cf. however \cite{o3}).  We note also the apparent denial of an entropy functional for
gravity without sources in \cite{c2}.  However in \cite{c2} an entropy is introduced via a
fluid stress energy tensor.  The theme of \cite{p1} does not seem to be threatened; the 
Einstein equations arise as a consistency condition indicating that spacetime structure 
(as defined by the Einstein equations!) is
robust under fluctuations.  The quantum potential in Section 2 ($Q=Q_G+Q_M$) arises via
the Hamiltonian context when one looks at a complex wave function $\psi=Aexp(iS/\hbar)$ with
$A$ and $S$ functionals of $h_{ij}$ and a matter field.  The interesting fact here is that 
$Q_G$ automatically arises once a complex (quantum) solution is sought (cf. (2.12)).  
The same feature arises in Section 4 where the introduction of a Bohmian context
corresponds to the entrance of quantum theory and the quantum potential automatically appears.
No matter potential is needed here and thus it seems that space time automatically contains
a quantum aspect which emerges when one looks at a Hamiltonian formulation with a complex
wave function (implicitly introducing a probability).  
The exact uncertainty approach of Section 3 introduces perturbations or
fluctuations in momentum based on fluctuations in $h_{ij}$ and exhibits the associated 
quantum potential.  The perturbations here are quite general in an explicit way and
essentially generate the amplitude of the wave function.  The form (3.7) in terms
of Fisher information automatically gives the fluctuations an entropic character (cf. 
\cite{o3} where one derives the SE on entropy ideas and see also Section 5.1
below); we will expand on this in Section 6.$\hfill\bs$

\subsection{WDW AGAIN}

We will rephrase some of this now following \cite{r5,r6} which clarifies the exact uncertainty
treatment of Sections 3-4.  We remark first that there seem to be strong relations between
the exact uncertainty method of deriving the SE and a technique of Olavo via entropy methods
(cf. \cite{c7,ch,o3}).  We sketch first the exact uncertainty method for the SE following
\cite{r5} (cf. also \cite{c1}).  For an ensemble of classical nonrelativistic particles
of mass $m$ moving in a potential V one has an ensemble Hamiltonian
\bq\label{a1}
\tl{H}_c[P,S]=\int dx\,P\left(\frac{|\na S|^2}{2m}+V\right)
\end{equation}
The ensuing equations of motion $(\dg)\,\,\pp_tP=(\gd\tl{H}_c/\gd S)$ and
$\pp_tS=-(\gd\tl{H}_c/\gd P)$ take the form
\bq\label{a2}
\pp_tP+\na\cdot\left(P\frac{\na S}{m}\right)=0;\,\,\pp_tS+\frac{|\na S|^2}{2m}+V=0
\end{equation}
Given stochastic perturbations $(*)\,\, p=\na S+f$ with 
$\overline{f}=0$ and $\overline{p}=\na S$ with $(\dg\dg)\,\,<E>=\int
dx\,P([\overline{|\na S+f|^2}/2m]+V)=\tl{H}_c+\int dx\,P\overline{f\cdot f}/2m$.
one asks for conditions on $f$ leading to quantum equations of motion and this is described
in \cite{c1} for example via four principles including exact uncertainty.  The quantum
ensemble Hamiltonian is then $(**)\,\,\tl{H}_q=\tl{H}_c+c\int dx(1/P)(|\na P|^2/2m)$
where the last term is a form of Fisher information.
\\[3mm]\indent
For more general situations following \cite{r6} one looks at the Fisher information matrix
\bq\label{a3}
I_{k\ell}=\int P(x')\left(\frac{\pp log(P(x^i))}{\pp x^k}\frac{\pp log(P(x^i))}{\pp x^{\ell}}
\right)d\mu(x^i)
\end{equation}
based on $y^i=\gt^i+x^i$ etc.  Using a standard 3-D metric $g_{ij}$ one obtains $(***)\,\,
\pp_tP+\sum g^{ik}\pp_i(P\pp-kS)=0$ from a variational principle via $(\dg\dg\dg)\,\,
\Phi_A=\int P(\pp_tS+(1/2)\sum g^{ik}\pp_iS\pp_kS)d^3xdt$ (varying S).  This leads trivially
to the classical HJ equation for a free particle upon variation in P so variation in S and P
leads to equations of motion for an ensemble of particles.  Now one can define the
information in P using the Fisher matrix via
\bq\label{a4} 
\Phi_B=\sum g^{ik}\int\frac{1}{P}\pp_iP\pp_kPd^3xdt=\sum g^{ik}I_{ik}
\end{equation}
Set then $\Phi=\Phi_A+\Phi_B$ and for variations of S and P vanishing at the boundary one
obtains
\bq\label{a5}
\pp_tP+\sum g^{ik}\pp_i(P\pp_kS)=0;\,\,\pp_tS+\sum (1/2)g^{ik}\pp_iS\pp_kS-
\end{equation}
$$-\gl\sum g^{ik}
\left(\frac{2}{P}\pp_i\pp_kP-\frac{1}{P^2}\pp_iP\pp_kP\right)=0$$
This is equivalent to the free particle SE
$(\bullet\bullet\bullet)\,\,i\hbar\pp_t\psi=\-(\hbar^2/2m)\sum g^{ik}\pp_i\pp_k\psi$
provided $\gl=\hbar^2/8$ and $\psi=P^{1/2}exp(iS/\hbar)$.  The connection to the quantum
potential comes here through (cf. (3.11))
\bq\label{a6}
\int PQd^3xdt=-\frac{\hbar^2}{8}\sum g^{ik}\int P\left(\frac{2}{P}\pp_i\pp_kP-\frac{1}{P^2}
\pp_iP\pp_kP\right)d^3xdt=
\end{equation}
$$=\frac{\hbar^2}{8}\sum g^{ik}\int\frac{1}{P}\pp_iP\pp_kPd^3xdt\sim \frac{\hbar^2}{8}
\Phi_B$$
which involves dropping a boundary term (note e.g. $\int_{\gO}\gD PdV=\int_{\pp\gO}\na P
\cdot{\bf n}d\gS$); this could become important in considerations of entropy and holography
(cf. Remark 1.3).
\\[3mm]\indent
Now going to \cite{r5} one considers
\bq\label{a7} 
H=\frac{1}{2}G_{ijk\ell}\frac{\gd kS}{\gd h_{ij}}\frac{\gd S}{\gd h_{k\ell}}-\sqrt{h}R=0;\,\,
H_i=-2D_j\left(h_{ik}\frac{\gd S}{\gd h_{kj}}\right)=0
\end{equation}
where $D_j$ is the covariant derivative and $(\bl\bl\bl)\,\,G_{ijk\ell}=(1/\sqrt{h})(h_{ik}
h_{j\ell}+h_{i\ell}h_{jk}-h_{ij}h_{k\ell})$ with $G=1/16\pi$
(this differs by a factor of 2 from a previous
$G_{ijk\ell}$).  As a consequence of the constraint $H=0$, \eqref{a7}, and $H_i=0$,
S must satisfy various constraints (including $\pp_tS=0$ - cf. \cite{b81}) and this
is all subsumed in the invariance of the HJ functional S under spatial coordinate
transformations.  Hence one can keep the Hamiltonian constraint, ignore the momentum
constraints, and reqire that S be invariant under the gauge group of spatial coordinate
transformations.  Now to define ensembles for gravitational fields one needs a measure
$Dh$ and a probability functional $P[h_{ij}]$ and this is discussed in some detail in
\cite{r5} following \cite{h81,m81} (we omit details here).  One is led to an ensemble
Hamiltonian and derived equations
\bq\label{a12}
\tl{H}_c=\int d^3x\int Dh\,PH;\,\,\pp_tP=\frac{\gD\tl{H}_c}{\gD
S};\,\,\pp_tS=-\frac{\gD\tl{H}_c}{\gD P}
\end{equation}
With $\pp_tS=\pp_tP=0$ the equations take the form
\bq\label{a13}
H=0;\,\,\int d^3x\frac{\gd}{\gd h_{ij}}\left(PG_{ijk\ell}\frac{\gd S}{\gd h_{k\ell}}\right)=0
\end{equation}
The latter equation corresponds to a continuity equation and in this spirit some argument
shows that it implies the standard rate equation
\bq\label{a15}
\pp_th_{ij}=NG_{ijk\ell}\frac{\gd S}{\gd h_{k\ell}}+D_iN_j+D_jN_i
\end{equation}
(as follows from the ADM formalism with N the lapse function and $N_j$ the shift vector).
Now writing $\pi^{k\ell}=(\gd S/\gd h_{k\ell})+f^{k\ell}$ with $\overline{f^{k\ell}}=0$ one
obtains an ensemble Hamiltonian (3.6) and an equation (3.7) as before.  Again putting
$\hbar=2\sqrt{c}$
and $\psi[h_{k\ell}]=\sqrt{P}exp(iS/\hbar)$ leads to the WDW equation (cf. (3.8))
\bq\label{a16}
\left[-\frac{\hbar^2}{2}\frac{\gd}{\gd h_{ij}}G_{ijk\ell}\frac{\gd}{\gd
h_{k\ell}}-\sqrt{h}R\right]\psi=0
\end{equation}
The procedures here suggest also replacing \eqref{a15} by
\bq\label{a17}
\pp_th_{ij}=NG_{ijk\ell}\left(\frac{\gd S}{\gd h_{k\ell}}+f^{k\ell}\right)+D_iN_j+D_jN_i
\end{equation}
Since the field momenta are subject to fluctuations so must be the extrinsic curvature
$K_{ij}=(1/2)G_{ijk\ell}(\gd S/\gd h_{k\ell})$ yielding then
\bq\label{a18}
K_{ij}=\frac{1}{2}G_{ijk\ell}\left(\frac{\gd S}{\gd h_{k\ell}}+f^{k\ell}\right)
\end{equation}

\section{THE ROLE OF THE QUANTUM POTENTIAL}
\renewcommand{\theequation}{6.\arabic{equation}}
\setcounter{equation}{0}

We will sketch here an approach to quantum gravity based on the quantum potential.
Generally one could start with WDW (which implies the Einstein equations with a few
assumptions (cf. \cite{g1}).  The introduction of a complex wave function $Aexp(iS/\hbar)$
automatically leads to a quantum potential ${\mf Q}$ and can be thought of as introducing a
statistical element into the picture via the amplitude A which should create a probability
density via $A^2=P$.  A Hamiltonian term arises then via $Q=\int {\mf Q}P$ which is 
proportional to a Fisher information and this specifies an ensemble of metric coefficients
$h_{ij}$ with probabilities $P[h_{ij}]$.  This does not involve a matter Lagrangian but
arises gratuitously from the metric term P since ${\mf Q}$ can be written entirely in terms
of P and the $h_{ij}$.  Thus the introduction of a complex wave function into a classical
problem is enough in itself to generate a quantum theory via information (or equivalently
entropy) ideas.
\\[3mm]\indent
The probability P can be thought of in various ways and the quantum potential term derived
from various points of view.  Thus in particular one can imagine momentum perturbations
$\pi^{ij}=(\gd S/\gd h_{ij})+f^{ij}$ as in $(\bgs)$ of Section 3 which via exact uncertainty
requires the $f^{ij}$ to be provided as in (3.7) producing a term
\bq\label{6.1}
\Phi_B\sim \frac{\hbar^2}{8}\int Dh\int dx NG_{ijk\ell}\frac{\gd P}{\gd h_{ij}}\frac
{\gd P}{\gd h_{k\ell}}
\end{equation}
We recall that the exact uncertainty theory for the SE involves a relation ${\bf (EU)}\,\,
\gd X\gD P_{nc}=\hbar/2$ where $P_{nc}$ with $<P_{nc}>=0$ is a nonclassical component of
momentum (cf. \cite{c1,h4,hal,r5}).  This is very similar in spirit to a formula of the form
$({\bf O})\,\,\overline{(\gd p)^2(\gd x)^2}=\hbar^2/4$ used by Olavo in deriving the SE
from entropy considerations (cf. \cite{ch,o3}).  The approach of Olavo is especially
interesting since it builds entropy explicitly into the theory and this could be identified
as its information content.
\\[3mm]\indent
In connection with the role of a complex wave function we recall that a complex velocity
has been emphasized by Castro, Mahecha, Nottale, and the author (cf.
\cite{c1,c17,c15,c16,n2}) and in a Weyl geometry with Weyl field $\phi_{\mu}\sim
A_{\mu}=-\pp_{\mu}log(P)$ (where $P\sim\rho$ is a density) a complex velocity $p_{\mu}+i\gl
A_{\mu}$ leads to
\bq\label{6.2}
|p_{\mu}+i\sqrt{\gl}A_{\mu}|^2=p_{\mu}^2+\gl A_{\mu}^2\sim g^{\mu\nu}\left(\frac{\pp S}{\pp
x^{\mu}}
\frac{\pp S}{\pp x^{\nu}}+\frac{\gl}{P^2}\frac{\pp P}{\pp x^{\mu}}\frac{\pp P}{\pp
x^{\nu}}\right)
\end{equation}
which generates again a Fisher matrix (cf. also \cite{w5}).
\\[3mm]\indent
The quantum potential also arises as a stress tensor in a quantum fluid (cf. \cite{c1,
dd,ta}) and in a diffusion context following \cite{c1,n3,n4} (cf. \cite{c1} for a survey
of the quantum potential).  There is also a quantum potential connected with a quantum matter
field $\phi$ as in Sections 2 and 4.  In such cases the nature of A as a probability or
density is less clear since it depends on $\phi$ and the $h_{ij}$.  One should be
able to form a Fisher information based on perturbations of both terms.  We see that the
quantum WDW equation is formed via a Fisher information type term in the Hamiltonian and in
view of the strong connection between entropy and Fisher information the idea of having an
entropy functional to extremize as in Section 1 seems eminently reasonable.

\end{document}